\documentclass[twocolumn, showpacs,preprintnumbers,amsmath,amssymb]{revtex4}
\usepackage{amssymb}
\usepackage{xcolor}
\usepackage{graphicx}% Include figure files
\usepackage{dcolumn}% Align table columns on decimal point
\usepackage{bm}% bold math
\usepackage{multirow}
\usepackage{booktabs}
\usepackage{natbib}
\usepackage{amssymb}
\usepackage{amsmath}	% Advanced maths commands
\usepackage{color}
\usepackage{cases} 

\begin{document}
\preprint{APS/123-QED}

\title{Dissociative recombination of CH$^+$ molecular ion induced by very low energy electrons}

\author{J. Zs. Mezei$^{1}$}%
\email[]{mezei.zsolt@atomki.hu}
\author{M. D. Ep\'ee Ep\'ee$^{2}$}%
\author{O. Motapon$^{2}$}
\author{I. F. Schneider$^{3,4}$}
\email[]{ioan.schneider@univ-lehavre.fr}
\affiliation{$^{1}$Institute for Nuclear Research (ATOMKI), H-4001 Debrecen, Hungary}%
\affiliation{$^{2}$Department of Physics, Faculty of Sciences, University of Douala, P. O. Box 24157, Douala, Cameroon}%
\affiliation{$^{3}$LOMC CNRS-UMR6294, Universit\'e le Havre Normandie, F-76058 Le Havre, France}%
\affiliation{$^{4}$LAC CNRS-FRE2038, Universit\'e Paris-Saclay, F-91405 Orsay, France}%
\date{\today}

\begin{abstract}
We used the multichannel quantum defect theory to compute cross sections and rate coefficients for the dissociative recombination of CH$^+$ initially in its lowest vibrational level $v_i^+ = 0$ with electrons of incident energy bellow $0.2$ eV. We have focused on the contribution of the $2$ $^2\Pi$ state which is the main dissociative recombination route at low collision energies. The final cross section is obtained by averaging the relevant initial rotational states $(N_i^+ = 0,\dots,10)$ with a $300$ K Boltzmann distribution.The Maxwell isotropic rate coefficients for dissociative recombination are also calculated for different initial rotational states and for electronic temperatures up to a few hundred Kelvins. Our results are compared to storage-ring measurements.
\end{abstract}

%\pacs{33.80. -b, 42.50. Hz}% PACS, the Physics and Astronomy
                             % Classification Scheme.
%\keywords{coupled-channel, optical shielding, KCs}%Use showkeys class option if keyword

                              %display desired
\maketitle

\section{Introduction}

The methylidyne cation CH$^{+}$ is one of the most important molecular species for astrophysical observations having a particular interest for the formation of large hydrocarbons in the interstellar medium (ISM). 
It was discovered at visible wavelengths by Douglas and Herzberg in 1941 \citep{Douglas41}, shortly after the discovery of the CH by Swings and Rosenfeld in 1937 \citep{Swings37}.

Since then, it has been detected in a wide variety of environments, but its ubiquity and high abundance in the diffuse environments in the ISM persists being a puzzling problem~\citep{valdivia2016}. The mechanism by which it forms has remained however elusive.

The most probable formation channel of the CH$^{+}$ cation, the reaction of atomic carbon C$^+$ with H$_2$, is endothermic by 0.398 eV (4620 K)~\citep{Hierl1997}, which is by far higher than the kinetic temperatures in ordinary diffuse clouds ($\sim 100$ K). 
In order to be formed in a sufficient abundance for observation, it has been suggested that the endothermicity related to its most probable formation pathway must be overcome by turbulent dissipation, shocks, or shears (see e.g.~\citep{valdivia2016}, and references therein). 
Furthermore, it was found that, in photon dominated regions, the ro-vibrationally excited H$_2$ reservoir can provide an alternative route to overcome the endothermicity of the formation reaction \citep{sternberg1995,agundez2010}, since the rotational and vibrational energies are as effective as the translational energy in promoting this type of reaction \citep{Hierl1997,zanchet2013}.

It is evident from the previous example that the detailed knowledge of the CH$^+$ molecular cation chemistry can provide unique physical insights into the modelling of the different ISM environments.
The full understanding of the production and loss mechanisms and the competition between the radiative processes, the destruction and collisional excitation processes are needed to be known in detail. 

 Due to its astrophysical interest the electron induced reactive processes of CH$^+$ has been extensively studied in the last three decades.
The first measurement of the dissociative recombination of CH$^+$ was performed by Mitchell and McGowan~\citep{Mm78}. A few years later,  Mul {\it et al.}~\citep{MulEtal81} reported measurements  which were slightly larger rate coefficients than those given by Mitchell and McGowan. Later on, more accurate values of thermal rate coefficients were given by Mitchell \citep{mitchell90}. In the mid 90s, the most comprehensive experimental study of CH$^+$ dissociative recombination was published by Amitay {\it et al.}~\citep{amitay96}, where cross sections, branching ratios and angular distributions have been reported. The special feature of these results was  the presence of prominent resonances that had not been present in the previous experimental and theoretical studies. They tentatively attributed these to capture of the incident electron into core excited Rydberg states, electronically coupled to the initial electronic continuum and the final dissociative channel.

The first theoretical results were reported by Takagi {\it et al.} \citep{tkd1991}. Their calculated cross sections agreed with the experimental results of Mitchell~\citep{mitchell90}. In order  to understand and characterise the broad resonances in the experiment of Amitay {\it et al.}~\citep{amitay96}, a Multichannel Quantum Defect (MQDT) calculation of DR was performed by Carata {\it et al.} \citep{carata2000} where they took into account the effect of core excited states. Their result have reproduced the broad resonances mentioned by Amitay {\it et al.}, but they have underestimated the experimental data by more than an order of magnitude, especially at higher collision energies. This was followed by the study of Guberman~\citep{guberman2005} on the angular distributions of the products of dissociative recombination of CH$^+$.
Recently, we have revised the work done by Carata {\it et al.} \citep{carata2000}, using an improved version of the MQDT method~\citep{kalyan2018}, and the newly calculated DR cross sections agree well with the results of Amitay {\it et al.}~\citep{amitay96}, at relatively high energy.

CH$^+$ is easily destroyed by reactions with electrons and hydrogen atoms, and also by reactions with H$_2$ molecules. The inelastic collisions with these dominant species, are no faster than the destructive reactive processes~\citep{faure2017}, thus, inelastic collisions can never fully equilibrate the rotational population of CH$^+$, leading to strong deviations of the level populations from local thermodynamic equilibrium.

In diffuse interstellar molecular clouds, 
where the temperature and pressure are very low, CH$^+$ mainly occurs in its ground vibrational state and, due to high electron densities, the main route for its destruction is via dissociative recombination (DR):
\begin{equation}
\label{eq:DR} \mbox{CH}^{+}(N_{i}^{+},v_{i}^{+}) + e^{-} \longrightarrow
(\mbox{CH}^{*},\mbox{CH}^{**}) \longrightarrow
\mbox{C}^{*} + \mbox{H}(1s)
\end{equation}

\noindent where $ N_{i}^{+} $ and $ v_{i}^{+} $ denote respectively, the initial rotational and vibrational levels of the ground-state of the molecular ion. %In the calculations we have taken 
The present work is devoted to give the theoretical and computational details on the rate coefficients reported in the paper~\citep{faure2017}, by presenting the methodology used in obtaining the cross sections of the dissociative recombination of CH$^+$, by taking into account in full detail the rotational effects.

The paper is structured as follows: section \ref{sec:theory} provides the main steps of our MQDT method. The computations of cross sections and rate coefficients and detailed comparison with former results are described in section ~\ref{sec:results}. The conclusions are presented in section ~\ref{sec:conclusions}. 

\section{Theoretical Method}\label{sec:theory}

In the present paper, we use an MQDT-type method to study the electron-impact collision processes given by eqs.~(\ref{eq:DR}) which result from the quantum interference of the {\it direct} mechanism - the capture takes place into a dissociative
state of the neutral system (CH$^{**}$) - and the {\it indirect} one -  the capture occurs \textit{via} a Rydberg state of the molecule CH$^{*}$ which is  predissociated by the CH$^{**}$ state. In both mechanisms the autoionization is in competition with the predissociation and leads, through the  reaction (\ref{eq:SEC}), 
 \begin{equation}
\label{eq:SEC} 
\mbox{CH}^{+}(v_{i}^{+}, N_{i}^{+}) +
e^{-}(\varepsilon) \longrightarrow
\mbox{CH}^{+}(v_{i}^{+}, N_{f}^{+}) +
e^{-}({\varepsilon}^{'})
\end{equation}
\noindent where $N_{i}^{+}$ and  $N_{f}^{+} $ are the initial and the final rotational quantum numbers of the molecular ion. In the calculations, for both reaction the initial vibrational quantum number was taken $ v_{i}^{+}=0$. The process is the so-called inelastic collision (IC), if $N_{i}^{+} <N_{f}^{+} $ and superelastic collision (SEC) for $N_{i}^{+} >N_{f}^{+}$.

\begin{figure}[t]
    \centering
    \includegraphics[width=1.0\columnwidth]{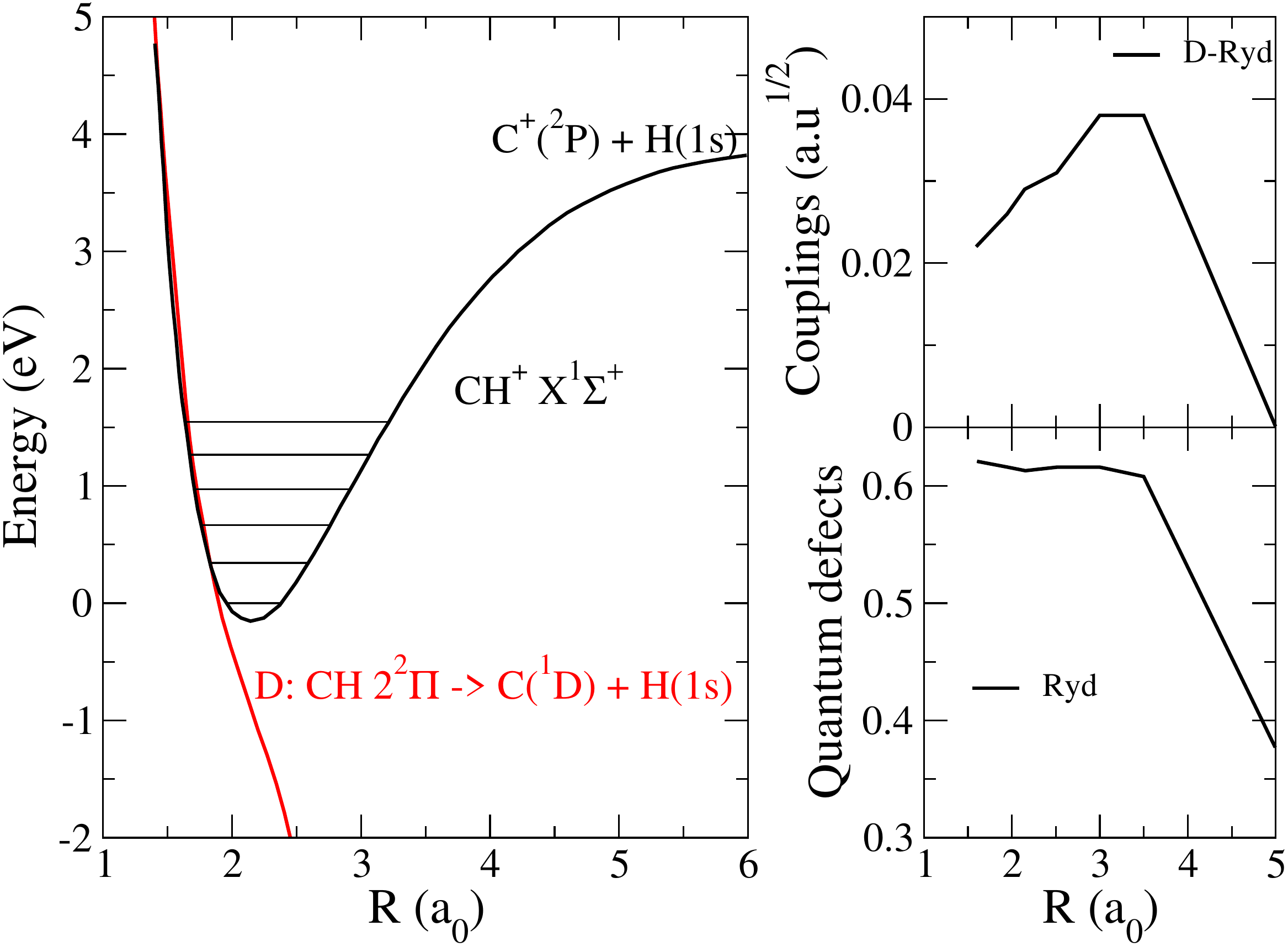}
    \caption{Molecular data sets relevant for the dissociative recombination of CH$^+$ compiled from Figs.~$2-5$ of Ref.~\citep{carata2000}. Left panel: the black line stand for the ground (X $^1\Sigma^+$) electronic state of the ion~\citep{green72}, having the C$^+(^2$P) $+$ H($1s$) dissociation limit, whereas the red continuous (D) line gives the dissociative autoionizing states (2 $^2\Pi$) of the neutral~\citep{tkd1991}. Upper right panel: electronic couplings: black line between the valence state (D) with the Rydberg states (Ryd) built on the ground ion state. Lower right panel: quantum defects for the Rydberg series based on the ground ion state.
} 
    \label{fig:courbe_pot}
\end{figure}

A detailed description of our theoretical approach is given in previous studies on H$_{2}^{+}$ and HD$^{+}$~\citep{annick1980,motapon2008,motapon2014,epee2015}.
The major steps of the method are briefly outlined as follows:
\begin{enumerate}
\item {\it Construction of the interaction matrix} $\boldsymbol{\mathcal{V}}$:\\
Within a quasi-diabatic representation~\citep{bardsley68a,bardsley68b,annick1980} of the CH states, the interaction matrix is based on the computed couplings between 
\textit{ionisation} channels - 
	associated to
	the ro-vibrational levels $N^+v^+$ of the cation  
	and 
	to the orbital quantum number $l$ of the incident/Rydberg electron - 
and 
\textit{dissociation} channels $d_j$.

\item {\it Computation of the reaction matrix} $\boldsymbol{\mathcal{K}}$:\\ 
Given $\boldsymbol{H_0}$ the Hamiltonian of the molecular system under study in which the Rydberg-valence interaction is neglected, we adopt  the second-order perturbative solution
for the Lippman-Schwinger integral equation \citep{ngassam03b}, 
written in operatorial form as:
\begin{equation}\label{eq:solveK}
\boldsymbol{\mathcal{K}}= \boldsymbol{\mathcal{V}} + \boldsymbol{\mathcal{V}}{\frac{1}{E-\boldsymbol{H_0}}}\boldsymbol{\mathcal{V}}.
\end{equation}
\noindent

\item {\it Diagonalization of the reaction matrix,} \\ 
yields the corresponding eigenvectors and eigenvalues which are used to build the eigenchannel wave functions.

\item {\it Frame transformation from the Born-Oppenheimer (short-range) to the close-coupling (long-range) representation,}\\ 
relying, for a given electronic total angular momentum quantum number $\Lambda$ and a given orbital quantum number of the incident/Rydberg electron $l$, on the vibrational wave functions of the molecular ion ($\chi_{N^{+}v^{+}}^{\Lambda ^{+}}$) and of the neutral system ($\chi_{Nv}^\Lambda$), on the quantum defect $\mu_{l}^{\Lambda}(R)$ and on the eigenvectors ($U_{lv,\alpha}^{\Lambda}$) and eigenvalues ($\eta_{\alpha}^{\Lambda}$) of the K-matrix:

\begin{align}
\label{eq:coeffCv}
{\cal C}_{lN^{+}v^{+}, \Lambda \alpha}&=\left(
\frac{2N^{+}+1}{2N+1}\right) ^{1/2}\left\langle l\left( \Lambda
-\Lambda ^{+}\right) N^{+}\Lambda ^{+}|lN^{+}N\Lambda
\right\rangle \nonumber \\
 &\times\frac{1+\tau^{+}\tau\left(-1 \right)^{N-l-N^{+}}} {\left[2\left(2-\delta_{\Lambda^{+},0} \right)\left(1+ \delta_{\Lambda^{+},0}\delta_{\Lambda,0} \right)   \right] ^{1/2}}\\
&\times\sum_{v} U_{lv,\alpha}^{\Lambda}\langle
\chi_{N^{+}v^{+}}^{\Lambda ^{+}}| \cos(\pi \mu_{l}^{\Lambda}
(R)+\eta_{\alpha}^{\Lambda})|\chi_{Nv}^{\Lambda}\rangle \nonumber
\end{align}

\noindent and

\begin{equation}
\label{eq:coeffCd}{\cal C}_{d_{j},\Lambda
\alpha}=U_{d_{j}\alpha}^{\Lambda}\cos \eta_{\alpha}^{\Lambda}
\end{equation}

\noindent as well as ${\cal S}_{lN^{+}v^{+},\Lambda \alpha }$ and
${\cal S}_{d_{j},\Lambda \alpha }$, which are obtained  by replacing cosine
with sine in Equations (\ref{eq:coeffCv}) and (\ref{eq:coeffCd}).

\item {\it Construction of the generalised scattering 
matrix $\boldsymbol{{X}}$}, \\ 
based on the frame-transformation coefficients, 
this matrix  being organised in blocks associated to open and/or closed ($o$ and/or $c$ respectively) channels:

\begin{equation}
\boldsymbol{{X}}=
 \left(\begin{array}{cc} \boldsymbol{X_{oo}} & \boldsymbol{X_{oc}}\\
                   \boldsymbol{X_{co}} & \boldsymbol{X_{cc}} \end{array} \right).
\end{equation}

\item {\it Construction of the generalised scattering matrix $\boldsymbol{\mathcal{S}}$}, \\ 
\begin{equation}\label{eq:solve3}
\boldsymbol{S}=\boldsymbol{X_{oo}}-\boldsymbol{X_{oc}}\frac{1}{\boldsymbol{X_{cc}}-\exp(-i2\pi\boldsymbol{ \nu})}\boldsymbol{X_{co}}.
\end{equation}
\noindent
based on the open channels, the first term in eq.~({\ref{eq:solve3}}), but also on their mixing with the closed ones, given by the second term, the denominator being responsible for the resonant patterns in the shape of the cross section
\citep{seaton83}. Here the matrix $\exp(-i2\pi\boldsymbol{ \nu})$ is diagonal and contains the effective quantum numbers $\nu_{v^{+}}$ associated to the vibrational thresholds of the closed ionisation channels.

\item {\it Computation of the cross-sections:}\\
For each of the relevant state of the neutral, characterised by the rotational quantum number $N$, which are grouped by symmetry properties: electronic total angular momentum quantum number $\Lambda$, electronic spin singlet/triplet, and for a given target cation ro-vibrational level $N^+_iv_i^+$ and energy of the incident electron $\varepsilon$, the dissociative recombination cross sections are computed as follows:

\begin{equation}\label{eqDR}
\sigma _{diss \leftarrow
N_{i}^{+}v_{i}^{+}}^{N}=\frac{\pi }{4\varepsilon
}\frac{2N+1}{2N_{i}^{+}+1}\rho\sum_{l,\Lambda,d_{j}}
|S^{{N\Lambda}}_{d_{j},l N_{i}^{+}v^{+}_{i}}|^{2},
\end{equation}
\noindent
where $\rho$ stands for the ratio between the multiplicity of the involved electronic states of CH and that of the target, CH$^+$.
\end{enumerate}

\section{Cross sections and rate coefficients}\label{sec:results}

The molecular data necessary to model the dissociative recombination and rotational
excitation are the potential energy curve (PEC) of the ground state of the ion, the PECs of the neutral valence dissociative states interacting with the ionization continua, those of the Rydberg states associated to these continua below the threshold (leading to smooth $R$-dependent quantum defects), and all the relevant Rydberg-valence couplings. These entities can be seen in figure~\ref{fig:courbe_pot}.

\begin{figure}[t]\centering
        \includegraphics[width=1.0\columnwidth]{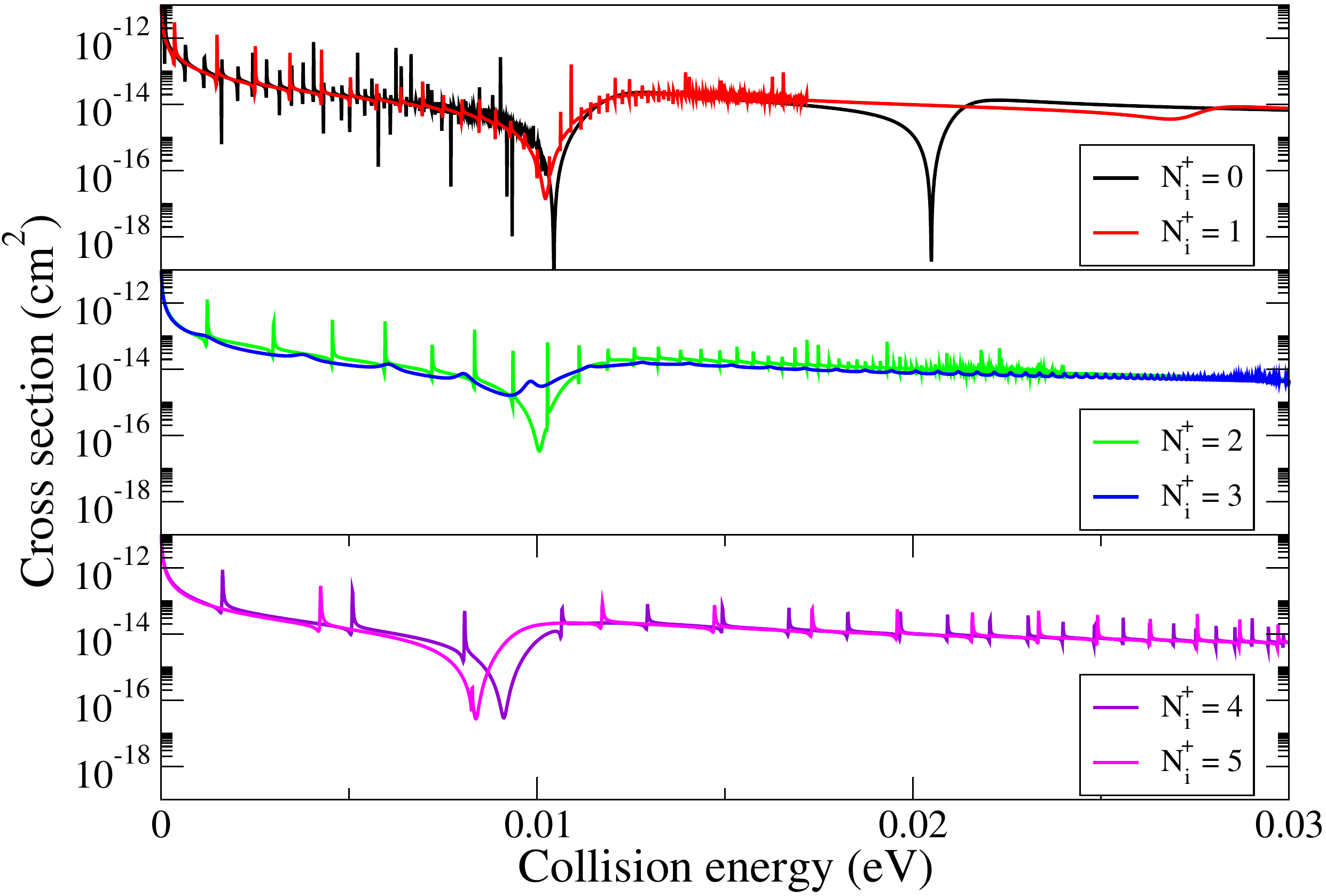}
    \caption{Cross sections for the dissociative recombination of vibrationally relaxed CH$^{+}(X^{1}\Sigma^{+})$ on initial rotational levels $N_{i}^{+}=0$ to $5$ for electron collision energyin the range $0-30$ meV.
    }
    \label{fig:cs_DR_1}
\end{figure}
\vspace{1cm}
\begin{figure}[h!]\centering
        \includegraphics[width=1.0\columnwidth]{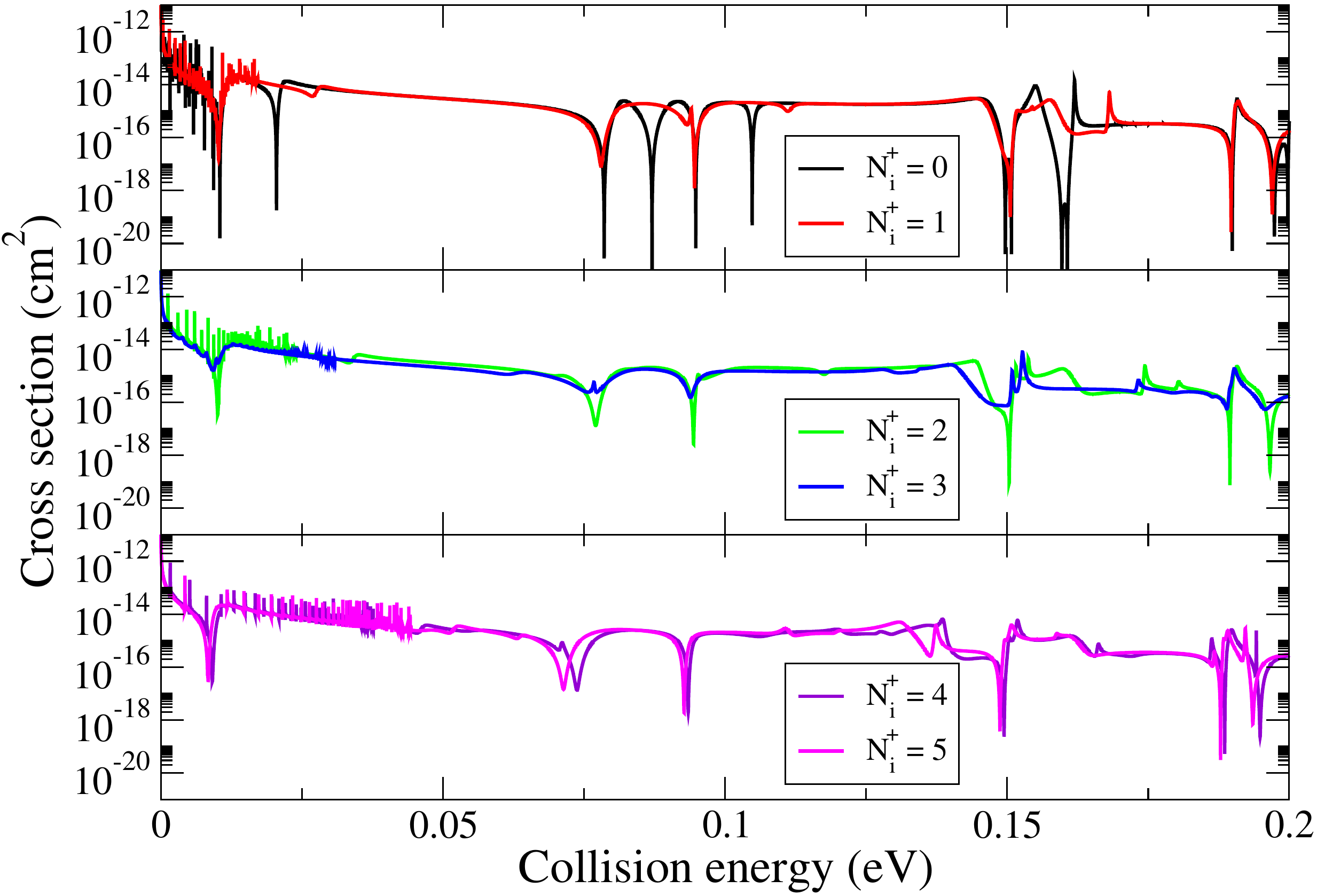}
   \caption{Cross sections for the dissociative recombination of vibrationally relaxed CH$^{+}(X^{1}\Sigma^{+})$ on initial rotational levels $N_{i}^{+}=0$ to $5$ for electron collision energy in the range $0-200$ meV.}
    \label{fig:cs_DR_2}
\end{figure}

In the present MQDT calculations, we used the potential curve of ground electronic state of CH$^{+}$ having $^1\Sigma^{+}$ symmetry that comprises in total 19 vibrational levels, calculated by Green {\it et al.} \citep{green72}, while for the dissociative doubly excited state (2$^{2}\Pi$) of the neutral molecule the PEC calculated by Takagi {\it
et al.} \citep{tkd1991}. All these PECs are shown in figure \ref{fig:courbe_pot} together with the first four vibrational levels of the ionic state. 
The $R$-dependent electronic coupling and quantum defect are those used by Carata {\it et al.} \citep{carata2000}, given in figures 3. and 4. of the same reference and are shown in the right part of fig.~\ref{fig:courbe_pot}.

Using this set of molecular data (PECs, electronic couplings and quantum defects), we performed a series of full rotational MQDT calculations of the cross sections for dissociative recombination for vibrationally relaxed CH$^{+}$ molecular ions on their first $11$ rotational levels ($v_{i}^{+}=0, N_{i}^{+}=0-10$), by taking into account the contribution of the $p$ ($l=1$) partial wave in representing the scattering.

  \begin{figure}[t]\centering
        \includegraphics[width=1.0\columnwidth]{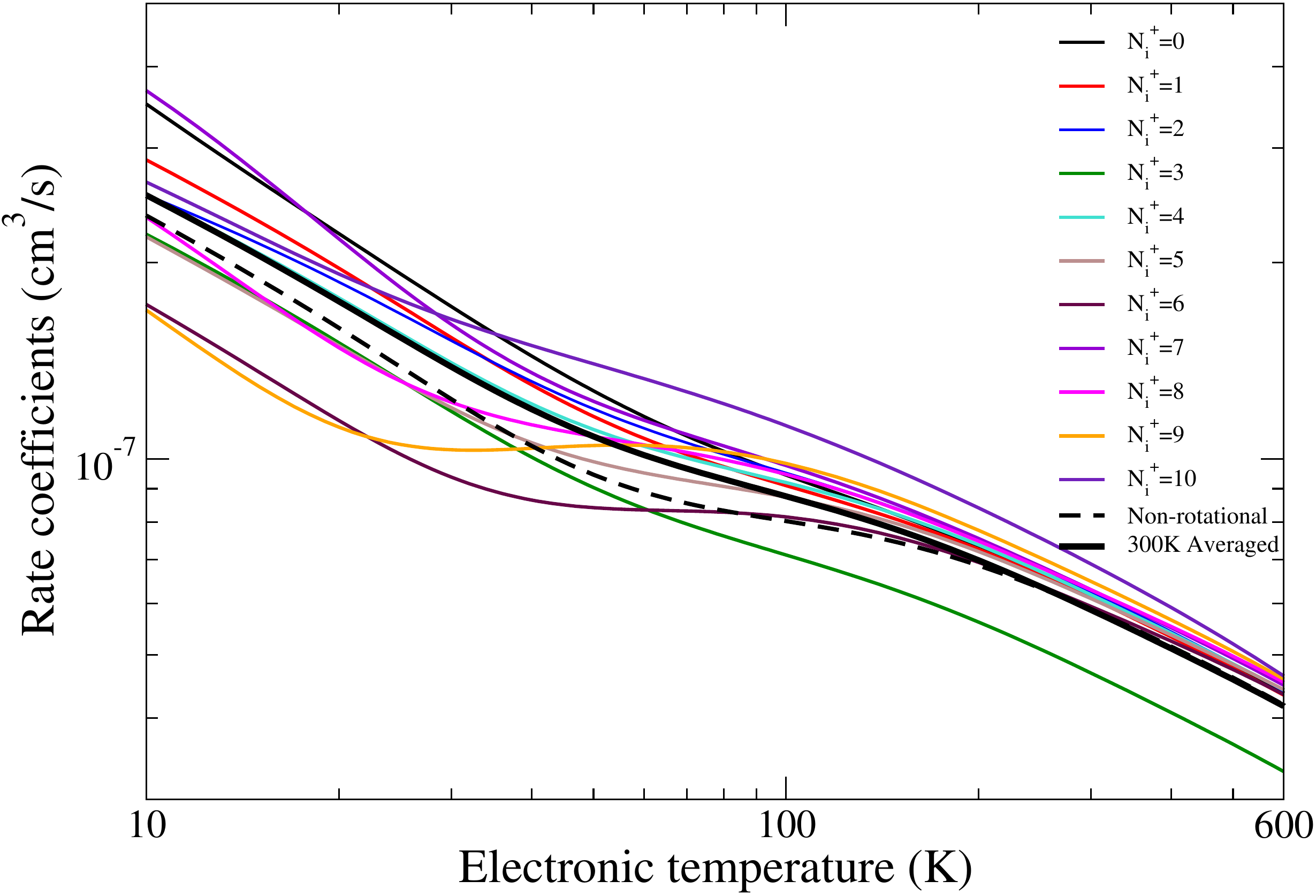}
   \caption{Maxwell isotropic rate coefficients for the
dissociative recombination
CH$^{+}(X^{2}\Sigma^{+})$ with $v_{i}^{+}=0$ as a function of
initial rotational level, $N_{i}^{+}=0$ to $10$.}
    \label{fig:rate_iso_DR}
\end{figure}
\vspace{1cm}
\begin{figure}[h!]\centering
         \includegraphics[width=1.0\columnwidth]{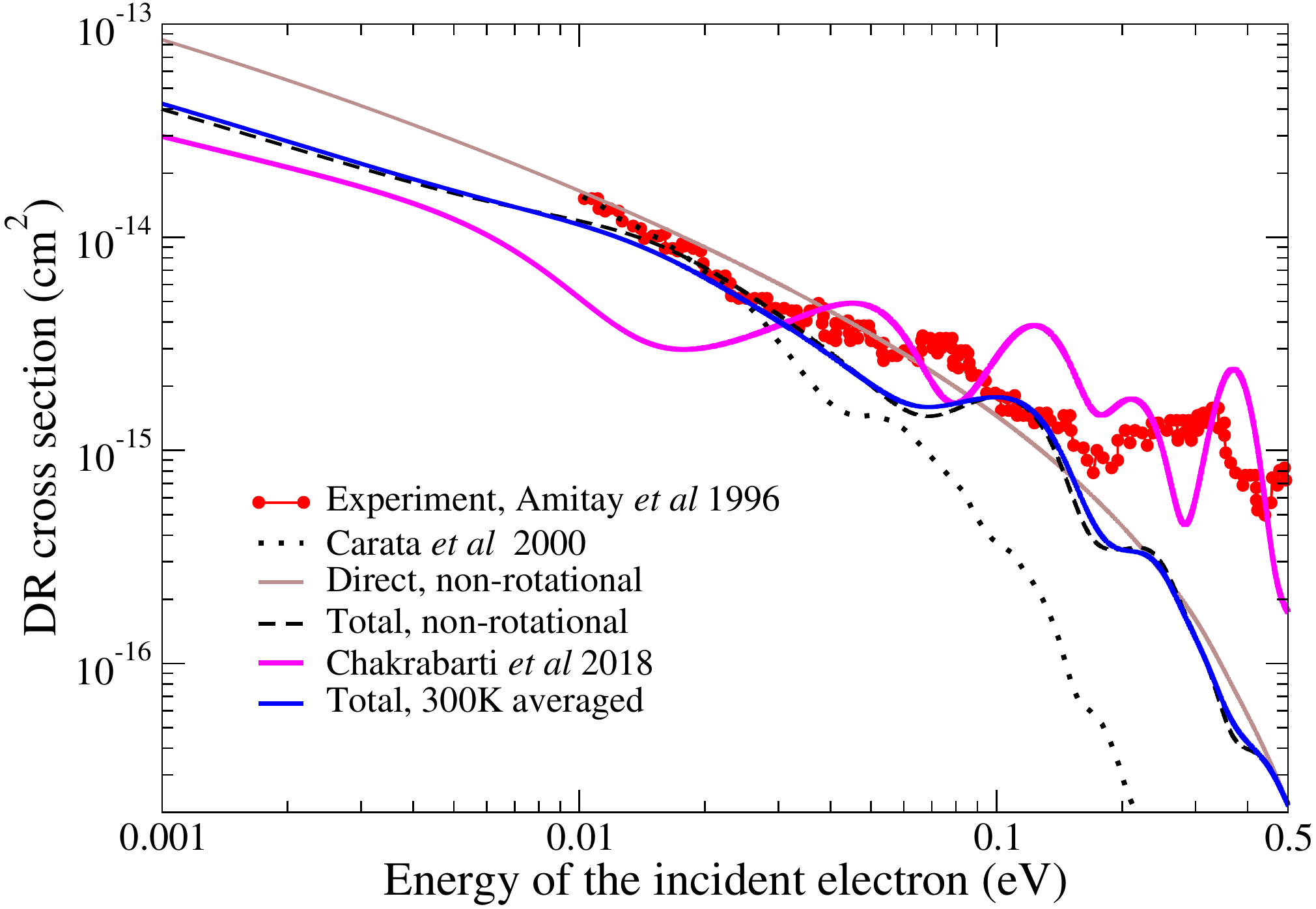}
\caption{Comparison of Maxwell anisotropic convoluted cross section for  the dissociative recombination of  CH$^{+}(X^{1}\Sigma^{+})$ in the ground vibrational state $ v_{i}^{+} =0$ with experimental measurement. Brown thin curve direct process without rotation,  black  dashed curve: total process without rotation, black solid curve: total process rotation included, black dotted line: MQDT calculation of Carata {\it et al.} \citep{carata2000}, red curve: experimental results of Amitay {\it et al.} \citep{amitay96}, thick magenta line: most recent three cores MQDT calculations without rotation~\citep{kalyan2018}.}
    \label{fig:cs_DR_ani_comp}
 \end{figure}
 
We have calculated the DR cross sections for the vibrationally relaxed ($v_{i}^{+}=0$) CH$^{+}$ molecular cation for all the rotational states up to $N_{i}^{+}=10$. In Figures \ref{fig:cs_DR_1} and \ref{fig:cs_DR_2} we show those for the first six rotational states  ($ N_{i}^{+}=0$ to $5$)
 for the energy range up to $30$ and $200$ meV respectively.

One may notice on one hand that the cross sections are governed by Rydberg resonances and on the other hand there is a clear dependence on the initial rotational state on the whole energy range, in both the absolute values of the cross sections and the intensities of the resonance structures. 

\noindent The DR Maxwell rate coefficients computed for the lowest $11$ rotational levels of the ground vibrational level, are represented in Figure \ref{fig:rate_iso_DR}, together with that one for non-rotational case and the averaged rate $300$ K. 
It can be noticed that there is a significant dependence on the initial rotational levels of the molecular cation for temperatures up to $200$ K. It is this region, where the difference between the averaged and non-rotational rates are the largest. Above this temperature, the dependence is less pronounced, the non-rotational rate coefficient and that one of averaged at $300$ K are on top of each other. 
\noindent  And finally, in Figure \ref{fig:cs_DR_ani_comp}, the convoluted DR cross section is compared to the non-rotational MQDT results and to the measurements. Our MQDT-based DR cross sections obtained including rotation show a satisfactory agreement with the experimental data up to 0.2 eV.  Above this collision energy the core excited effects become important.
We also present in  this figure, the results of Carata {\it et al.} \citep{carata2000}, obtained using an MQDT first order perturbative treatment of dissociative recombination with the inclusion of the first two core excited states of the ion (black dotted line), whereas  our most recent results were calculated with the ground ionic state but in second order (thick magenta line). It can be seen in the energy range considered that the rotational effects and the second order treatment brings remarkable enhancement in the DR cross sections.  
 
%%%%%%%%%%%%%%%%%%%%%%%%%%%%%%%%%%%%%%%%%%
\section{Conclusions}\label{sec:conclusions}

In the present work, we have computed  cross section and rate coefficients for dissociative recombination of CH$^{+}$
in its  ground vibrational state $ v_{i}^{+} $=0. We show the dependence of the rate coefficients and cross section on initial rotational quantum numbers. Whereas the agreement between our results and the experimental results is not perfect, there are no alarming discrepancies either. The cross section results for DR  bellow $0.2$ eV
as well as the rate coefficients for temperatures smaller than $600$ K as function of the initial rotational quantum numbers presented here can be used for astrophysical models.

In order to extend our model to rotational transitions and allow comparisons with the the R-matrix method combined with the adiabatic nuclei-rotation approximation calculations~\citep{hamilton2015}, as well as to improve our approach valid for high energy~\citep{kalyan2018}, we have to include further molecular states starting from the most recent calculations of Chakrabarti {\it et al}~\citep{kalyan2019}, and to take into account further partial waves and excited core effects.  

\section*{Acknowledgements}
%%%%%%%%%%%%%%%%%%%%%%%%%%%%%%%%%%%%%%%%%%%%%%%%%%
JZsM is grateful for the support of the National Research, Development and Innovation Fund of Hungary, under the K18 funding scheme with project no. K 128621. IFS  acknowledges support from Agence Nationale de la Recherche via the project MONA, from the CNRS via the  GdR  TheMS, from La R\'egion Normandie, FEDER and LabEx EMC$^3$ via  the projects Bioengine, PicoLIBS, EMoPlaF and CO$_2$-VIRIDIS, and from the Programme National "Physique et Chimie du Milieu Interstellaire" (PCMI) of CNRS/INSU with INC/INP co-funded by CEA and CNES.
%%%%%%%%%%%%%%%%%%%% REFERENCES %%%%%%%%%%%%%%%%%%
\section*{The author contributions}{All authors were equally involved in the calculations reported in the present paper as well as in the writing of the manuscript.}

%%%%%%%%%%%%%%%%%%%%%%%%%%%%%%%%%%%%%%%%%%
\section*{Funding}{This research received no external funding.}
\section*{Conflicts of interest}{The authors declare no conflict of interest.}
%\section*{Data availability}
%The data underlying this article will be shared on reasonable request to the corresponding author.

\end{document}